\documentclass[aps,prb,twocolumn,showpacs,reprint,superscriptaddress,longbibliography]{revtex4-1}
\usepackage{amsmath, amsfonts, amssymb}
\usepackage{enumerate}
\usepackage{graphicx}
\usepackage{float}
\usepackage{color}
\usepackage[breaklinks]{hyperref}
\hypersetup{colorlinks,linkcolor={magenta},citecolor={blue},urlcolor={blue}}

\usepackage{paralist}
\usepackage{multirow}

\newcommand{\eref}[1]{Eq.~(\ref{#1})}
\newcommand{\fref}[1]{Fig.~\ref{#1}}

\newcommand{\mean}[1]{\langle #1 \rangle}

\begin{document}

\title{Dirac point formation revealed by Andreev tunneling in superlattice-graphene/superconductor junctions}
\author{Shirley G\'omez P\'aez}
\affiliation{Departamento de F\'{\i}sica, Universidad Nacional de Colombia, Bogot\'a,
Colombia}
\affiliation{Departamento de F\'{\i}sica, Universidad el Bosque, Bogot\'a, Colombia}
\author{Pablo Burset}
\affiliation{Department of Applied Physics, Aalto University, 00076 Aalto, Finland}
\author{Camilo Mart\'{\i}nez}
\affiliation{Departamento de F\'{\i}sica, Universidad Nacional de Colombia, Bogot\'a,
Colombia}

\author{William J. Herrera}
\affiliation{Departamento de F\'{\i}sica, Universidad Nacional de Colombia, Bogot\'a,
Colombia}
\author{Alfredo Levy Yeyati}
\affiliation{Departamento de F\'{\i}sica Te\'orica de la Materia Condensada and Condensed Matter Physics Center (IFIMAC) and Instituto Nicol\'as Cabrera, Universidad Aut\'onoma de Madrid, E-28049 Madrid, Spain}

\begin{abstract}
A graphene superlattice is formed by a one-dimensional periodic potential and is characterized by the emergence of new Dirac points in the electronic structure.
The group velocity of graphene's massless Dirac fermions at the new points is drastically reduced, resulting in a measurable effect in the conductance spectroscopy.
We show here that tunnel spectroscopy using a superconducting hybrid junction is more sensitive to the formation of Dirac points in the spectrum of graphene superlattices due to the additional contribution of Andreev processes.
We examine the transport properties of a graphene-based superlattice--superconductor hybrid junction and demonstrate that a superlattice potential can coexist with proximity-induced superconducting correlations. Both effects contribute to change graphene's spectrum for subgap energies and, as a result, the normalized tunneling conductance features sharp changes for voltages proportional to the energy separation between the original and the newly generated Dirac points.
Consequently, the superconducting differential conductance provides an excellent tool to reveal how the new Dirac points emerge from the original band. This result is robust against asymmetries and finite-size effects in the superlattice potential and is improved by an effective doping comparable to the superconducting gap.
\end{abstract}

\maketitle

\section{Introduction. }

\begin{figure}[b]
\centering\includegraphics[width=1.0\columnwidth]{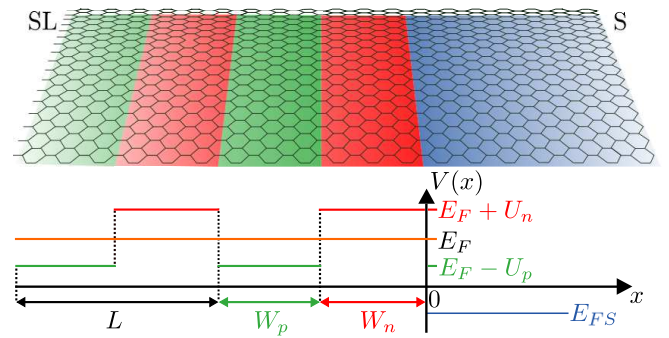}
\caption{Semi-infinite superlattice (SL) coupled to a superconductor (S) on top of a graphene layer, with an sketch of the system's energy diagram. }
\label{fig:potencial}
\end{figure}

Graphene is a versatile material that can be modified to be metallic or semiconducting, owing to its gapless, linear low-energy spectrum~\cite{Sarma_RMP,Nori_PhysRep,DiracMaterials}.
This duality can be exploited in graphene superlattices, formed by a periodic, one-dimensional electrostatic potential on the graphene sheet, making graphene a promising candidate for designed electronic circuits.
It is theoretically established that charge carriers in graphene behave like chiral Dirac fermions. Under a one-dimensional superlattice potential of amplitude $U$ and period $L$, see \fref{fig:potencial}, chirality forbids the opening of a band gap, instead creating new Dirac points (DPs) when the product  $U\cdotp L$ reaches a critical value~\cite{Brey_2009,Barbier_2010,Burset_2011}.

The propagation of chiral Dirac fermions under superlattices is highly anisotropic and can be controlled by varying the superlattice potential and Fermi energy. The resulting carrier velocity can be completely suppressed along one direction, but remains barely unchanged in the opposite, allowing for collimated electron beams~\cite{Park_2008}.
Experimental realization of high-quality periodic superlattice potentials on graphene has been achieved using boron nitride encapsulation~\cite{Decker_2011,Xue_2011}. Dirac point formation was measured as resistivity peaks~\cite{Yankowitz_2012,Ponomarenko_2013,Lee_2016}, paving the way for novel and exotic physics in graphene-based superlattices~\cite{Cao_2018,Cao_2018b}.

The recent discovery of unconventional superconductivity in bilayer graphene superlattices~\cite{Jarillo_2018} has pointed to the interesting connection between superconducting correlations and superlattice potentials in graphene.
However, to the best of our knowledge, the interplay between superconductivity and Dirac point formation by one-dimensional superlattices has not been explored yet.
Graphene--superconductor hybrids can now be fabricated in high-quality transparent junctions that work in the ballistic regime \cite{Schonenberger_2012,Klapwijk_2015,Geim_2016,Review_Graphene-SC}.
In such hybrid junctions, electrons and holes from the conduction band of the normal lead combine to form Cooper pairs in the superconductor by means of a microscopic process known as Andreev reflection~\cite{Beenakker_RMP}.
When the doping is smaller than the applied voltage and the superconducting gap, graphene's peculiar gapless dispersion allows for an unusual Andreev reflection where conduction band electrons are converted into holes belonging to the valence band~\cite{Specular}.
Here, we demonstrate that these inter-band Andreev processes provide unique signatures of the formation of Dirac points, which could substantially facilitate their experimental detection.

Advances in experimental control of graphene devices are leading to a series of remarkable works reporting inter-band Andreev reflections\cite{Efetov_2016}, spectroscopy of Andreev bound states in Josephson junctions~\cite{Dirks_2011}, splitting of Cooper pairs~\cite{Hakonen_2015}, and proximity-induced unconventional superconductivity~\cite{Tonnoir_2013,Ludbrook_2015,Chapman_2016,Hyoyoung_2017,Robinson_2017}.
Since a superlattice potential can be seen as a series of p-n junctions, cf. \fref{fig:potencial}, an example of the potential applications of superlattice--superconductor hybrids is the recently measured~\cite{Lee_2015} focusing of beams of Dirac fermions in graphene-based p-n junctions~\cite{Cheianov_2007,Cserti_2007,Sun_2010}.
Existing theoretical works extend such effects to graphene junctions involving ferromagnets~\cite{Zareyan_2010} and superconductors~\cite{Gomez_2012}.

In this paper, we analyze the interplay between a superlattice potential and proximity-induced superconductivity on graphene.
We focus on the effect of the emergence of new DPs on the transport properties of a graphene-based superlattice--superconductor (SL-S) junction.
We demonstrate that subgap transport is extremely sensitive to the creation of new DPs when inter-band Andreev processes are dominant.
Strikingly, the differential conductance presents sharp changes for voltages proportional to the energy separation between the original and the newly generated DPs. This effect is robust against asymmetries in the superlattice potential, the presence of an additional doping on the graphene layer, and finite-size effects. Therefore, graphene-based SL-S junctions are a convenient setup for addressing fundamental questions about the formation of new chiral Dirac fermions by periodic potentials and a promising component of future graphene-based electronic circuits with tailored properties.



\section{Differential conductance of a graphene-based superlattice-superconductor junction}


We consider an infinite graphene plane where a superlattice potential $V(x)$ is created on the $x<0$ region and superconductivity is induced by proximity effect on the half-plane with $x>0$. The resulting graphene-based SL-S junction is sketched in \fref{fig:potencial}. Assuming valley and spin degeneracy~\cite{Specular,Herrera_2010,Oscar_2018}, the Dirac-Bogoliubov-de Gennes (DBdG) Hamiltonian in sublattice and particle-hole (Nambu) spaces is given by
\begin{equation}
\check{H}_\text{DBdG}=\left(
\begin{array}{cc}
\hat{h}+V(x)\hat{\sigma}_{0} & \hat{\Delta}(x) \\
\hat{\Delta}^{\dag }(x) & -\hat{h}-V(x)\hat{\sigma}_{0}%
\end{array}%
\right)  , \label{DBdG}
\end{equation}
with $\hat{h}=v_{F}(-i\partial_{x}\hat{\sigma}_{1}+q\hat{\sigma}_{2})-\epsilon\hat{\sigma}_{0}$ the single-particle Dirac Hamiltonian, $q$ the conserved component of the wave vector parallel to the interface, $v_{F}$ the Fermi velocity, and $\hat{\sigma}_{0,1,2,3}$ the Pauli matrices in sublattice space.
The superlattice potential consists of a periodic repetition of potential barriers and wells with height $\pm U_{n,p}$ and width $W_{n,p}$, cf. \fref{fig:potencial}. Explicitly,
\begin{equation}\label{eq:potencial}
V(x)=\left\{
\begin{array}{cl}
E_F+U_n, & x\in \left[ mL,mL-W_{n}\right]  \\
E_F-U_p, & x\in \left[mL-W_{n},\left( m-1\right)L\right]
\end{array} \right. ,
\end{equation}
with $m\!=\!0,-1,-2,\dots$, and $L\!=\!W_{n}+W_{p}$ the period of the superlattice potential.

We denote the Fermi energy in the superconducting regions as $E_{FS}$.
We consider rigid boundary conditions for a conventional proximity-induced $s$-wave pairing $\hat{\Delta}(x)\!=\!\Delta \hat{\sigma}_{0}\Theta (x)$, with $\Delta\! >\!0$ the pairing amplitude and $\Theta (x)$ the Heaviside step
function.
Approximating the spatial dependence of the pairing potential by a step
function is valid as long as the Fermi wavelength of the quasi-particles in
the superconductor is much smaller than the superconducting coherence
length $\xi_0\!=\!\hbar v_F/\Delta$, i.e., $E_{FS}\!\gg\! \Delta$.

\begin{figure*}
\includegraphics[width=1.\textwidth]{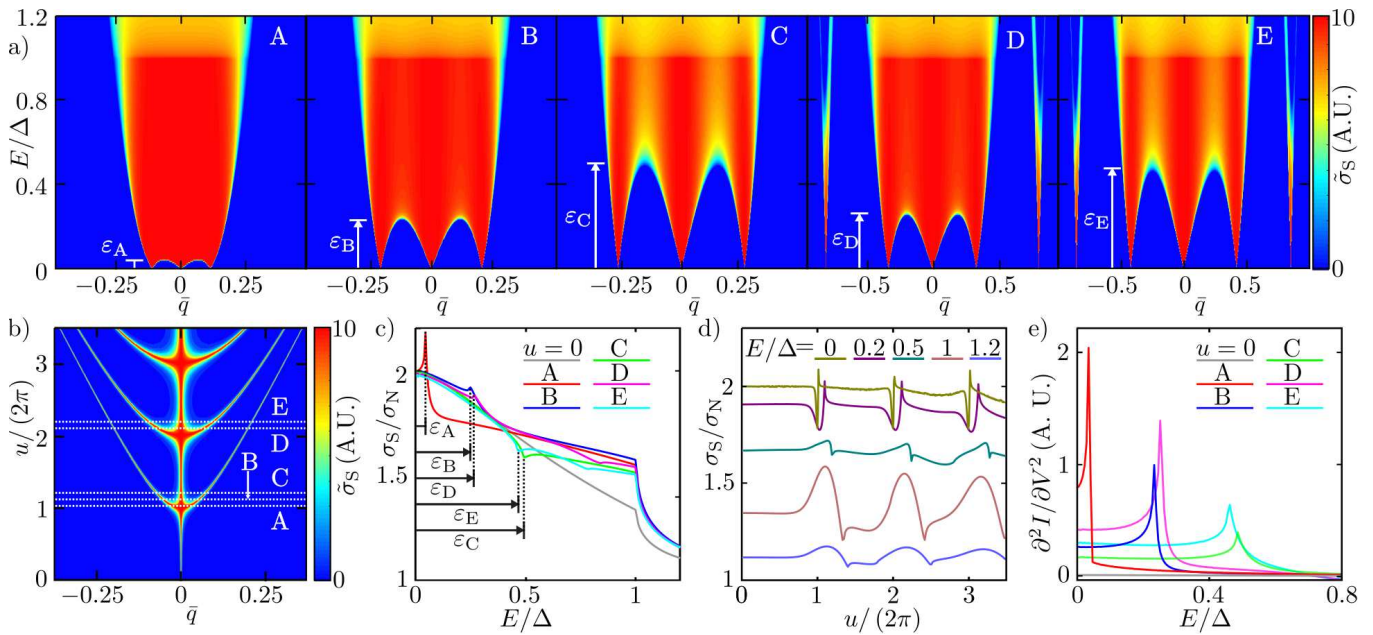}
\caption{Undoped, symmetric graphene superlattice.
a) Spectral differential conductance for different values of the superlattice potential $u/\left(2\pi\right)\!=\!1.03$(A), $1.11$(B), $1.19$(C), $2.15$(D), $2.23$(E).
b) Map of the zero-energy spectral differential conductance as a function of $u$ and $\bar{q}\!=\!q/q_\text{max}$, with $\hbar v_F q_\text{max}\!\simeq\!29\Delta$.
c) Differential conductance as a function of the energy for the different values of $u$ used in a). The parameter $\protect\varepsilon $ indicates the separation of the new pair of Dirac points from the original cone.
d) Differential conductance as a function of $u$ for different energies.
e) Estimation of $\protect\varepsilon $ from the second derivative of the current with respect to the voltage. For all cases $E_{F}\!=\!0$ and $L\!=\!\protect\xi _{0}/2$.
}
\label{fig:sym}
\end{figure*}

The transport properties of the SL-S junction are encoded in the retarded Green function $\check{g}_{q}^{r}(x,x^\prime )\!=\!\int \mathrm{d}q e^{iq(y-y^{\prime })} \check{g}_{q}^{r}(x,x^\prime ,y,y^{\prime })$, which satisfies the non-homogeneous DBdG equation
\begin{equation}\label{DBdG-GF}
\left[ \left( E+i0^{+}\right) \check{I}-\check{H}_\text{DBdG}\right] \check{g}_{q}^{r}(x,x^\prime )=\delta (x-x^\prime )\check{I} ,
\end{equation}
with $\check{I}$ the identity matrix.
A solution of \eref{DBdG-GF} is obtained combining asymptotic solutions of \eref{DBdG} that obey the boundary conditions at the edges of a finite length graphene sheet, following a generalization of the method developed in Refs.~\onlinecite{McMillan_1968,Furusaki_1991,Kashiwaya_RPP,Herrera_2010,Lu_2015,Crepin_2015,Burset_2015,Keidel_2018,Lu_2018,Breunig_2018}.
The Green function for the SL can then be written as
\begin{equation}\label{eq:gf_SL}
\hat{g}_{\text{SL}}(0^-,0^-)=\frac{i}{2\hbar v_{F} C_{n}}\left(
\begin{array}{cc}
0 & 2 C_{n} \\
0 &  M+\sqrt{J+M^{2}}
\end{array}%
\right) ,
\end{equation}%
with
\begin{gather*}
M= C_{p}C_{n}-D_{pn}^{2} -1 , \quad J=4 C_{n}C_{p} , \\
C_{p(n)}=\frac{ c_{p(n)}\left( c_{n(p)}^{2}+d_{n(p)}^{2}\right)
+c_{n(p)}}{c_{p}c_{n} + 1}, \,
D_{pn}= \frac{i d_{p}d_{n}}{c_{p}c_{n} + 1},
\end{gather*}
and
\begin{align*}
c_{n(p)}={}& \frac{e^{-i\alpha \lbrack \epsilon(\chi)]}\left(
e^{2ik_{x}[\epsilon(\chi)]W_{n(p)}}-1\right) }{1+e^{-2i\alpha \lbrack
\epsilon(\chi)]}e^{2ik_{x}[\epsilon(\chi)]W_{n(p)}}}, \\
d_{n(p)}={}& \frac{e^{-ik_{x}[\epsilon(\chi)]W_{n(p)}}\left( 1+e^{-2i\alpha
\lbrack \epsilon(\chi)]}\right) }{1+e^{-2i\alpha \lbrack
\epsilon(\chi)]}e^{-2ik_{x}[\epsilon(\chi)]W_{p(n)}}}.
\end{align*}
Here, $e^{\pm i\alpha_{e(h)} \lbrack \epsilon]}\!=\!\hbar v_{F}(k^{e(h)}_{x}[\epsilon]\!\pm iq)/(\epsilon \pm E)$ and $k^{e(h)}_{x}[\epsilon]\!=\!\sqrt{[ (\epsilon +E)/\hbar v_{F}] ^{2}-q^{2} }$, with $E$ and $\epsilon$ the excitation and potential energies, respectively.
We consider a transparent coupling between the superlattice and the superconductor.
For more details of the calculations, we refer the reader to Appendix~\ref{sec:app1}.

The differential conductance depends on the potential difference between SL and S, $V\!=\!V_\text{SL}-V_\text{S}$, and can be written as
\begin{equation}\label{conductance}
\sigma_\text{S} (V)=\frac{\partial I}{\partial V}=\sigma _\text{Q}(V)+\sigma _\text{A}(V),
\end{equation}
where $\sigma _\text{A}$ ($\sigma _\text{Q}$) represents the contribution of the Andreev (quasiparticle) processes.
Here,
\begin{equation*}
\sigma _\text{Q(A)}(E)= 8\pi ^{2}\frac{e^{2}}{h}\int \mathrm{d}q\tilde{\sigma}_\text{Q(A)}(E,q),
\end{equation*}
with $\tilde{\sigma}_\text{Q(A)}$ defined in terms of the Green functions in Appendix~\ref{sec:app2}. We normalize our results using the conductance for $\Delta=0$, $\sigma_\text{N}$.

\section{Ideal superlattice}

We analyze the transport properties of a SL-S junction to illustrate how the emergence of Dirac points by the superlattice potential is neatly captured in the subgap differential conductance.
We start considering an \textit{ideal} superlattice potential, i.e., semi-infinite, created around the charge neutrality point, $E_F\!=\!0$, and symmetric with $U_{p}\!=\!U_{n}\!\equiv\!U$ and $W_p\!=\!W_n\!=\!L/2$, see \fref{fig:potencial}.
The normalized barrier strength is thus given by $u\!=\!UL/\hbar v_{F}$. Under these conditions, a new set of DPs is created when $u\!=\!2n\pi$, with $n$ a positive integer~\cite{Brey_2009,Barbier_2010,Burset_2011}.

As shown in \fref{fig:sym}(a,b), the superlattice potential can coexist with proximity-induced superconducting pairing in the superlattice region close to the interface with the superconductor.
Indeed, the energy-momentum plots in \fref{fig:sym}(a), calculated for several values of the superlattice strength, clearly show how the band dispersion relation for subgap energies is modified after the first (panels A, B, C) and second (D, E) generation of DPs.
The condition for the formation of DPs is thus not altered by the presence of the superconductor. Setting $E\!=\!0$, we plot in \fref{fig:sym}(b) the spectral conductance $\sigma(E\!=\!0,q)$ as a function of the transverse momentum $\hbar q$ and the barrier strength $u$. The first pair of DPs is formed at the critical value $u\!=\!2\pi$, and the second when $u\!=\!4\pi$ is reached. The spectral density $\sigma(E,q)$ has been calculated in the SL region close to the interface with the superconductor.
We identify the \textit{menorah}-like structure of \fref{fig:sym}(b) as the fingerprint of the superlattice.

In the absence of the superlattice potential, a small trace of the original Dirac cone centered around $q\!=\!0$ remains for energies below the superconducting gap $\Delta$, cf. Refs.~\onlinecite{Burset_2008,Burset_2009,Burset_2013b}.
The superlattice potential changes the Fermi velocity, thus widening graphene's cone-like spectrum, even before creating new DPs.
Importantly, when coupled to a superconductor, the spectral density of states for subgap energies is higher compared to the non-superconducting case with $\Delta\!=\!0$, which can be seen as a sharp change in color for energies below and above the gap in \fref{fig:sym}(a).
Finally, as $u$ increases beyond the critical value [panels A-C in \fref{fig:sym}(a)], a new pair of Dirac points appear.
The original DP and the newly created ones are disconnected at zero energy, but they merge for finite energies over a characteristic value $\varepsilon$, indicated by white arrows in \fref{fig:sym}(a).
The parameter $\varepsilon$ becomes finite for $u\!>\!2\pi$ and increases with $u$ until it approaches $\Delta$ as a new pair of DPs is completely formed when $u\!=\!4\pi$. At this critical value, $\varepsilon$ is set to zero, and becomes finite again when $u\!>\!4\pi$. The energy parameter $\varepsilon$ thus provides a qualitative measure of the separation between the original DP and every new pair.

These three effects, namely,
\begin{inparaenum}[(i)]
    \item the formation of DPs when $u\!=\!2n\pi$;
    \item an enhanced spectral density for subgap energies; and
    \item the separation between DPs characterized by the parameter $\varepsilon$,
\end{inparaenum} can be neatly observed in the normalized differential conductance $\sigma_\text{S}/\sigma_\text{N}$.
\fref{fig:sym}(c) and (d) show the normalized differential conductance of the SL-S junction as a function of the excitation energy and superlattice strength, respectively.
In the absence of superlattice potential, since $E_F\!=\!0$, we recover the conductance results of Ref.~\onlinecite{Specular} where inter-band Andreev reflection is dominant (gray line with $u\!=\!0$).
For $u\!\neq\!0$, the normalized conductance at zero energy remains fixed at $2\sigma_0$, except at the critical values where DPs are formed, see \fref{fig:sym}(d). Since the normal state conductance $\sigma_0$ accounts for the contribution from three channels, the doubling of the conductance is due to perfect Andreev reflection taking place at the original and the new DPs.

By contrast, for finite energies we observe two effects connected to the presence of a superlattice potential.
First, the conductance features a peak or a dip for energies equal to the parameter $\varepsilon$ [see arrows in \fref{fig:sym}(c)].
Second, the superlattice increases the conductance for energies close to the gap ($E\!\lesssim\!\Delta$). The latter is a direct consequence of the superlattice-enhanced spectral density for subgap energies. The former can be associated to the different contribution from Andreev and normal reflections.
For small values of $\varepsilon$, the normalized conductance exhibits a peak indicating that inter-band Andreev processes become dominant when the DPs are created. As $\varepsilon$ approaches $\Delta$, the peak in the conductance becomes a dip since the DPs are more clearly separated thus producing more backscattering for $q$ values between DPs.
In both cases, the parameter $\varepsilon$, which indicates the separation between the newly created DPs and the original band, is accessible through the differential conductance. By taking the derivative of the conductance with respect to the voltage (i.e., $\mathrm{d}^2I/\mathrm{d}V^2$), the small kinks in the conductance shown in \fref{fig:sym}(c) appear as clear peaks in the derivative, cf. \fref{fig:sym}(e).

The above results show that a superlattice potential can create new DPs in the presence of superconducting correlations.
As long as the strength of the potential is comparable to the superconducting gap, the zero-energy normalized conductance is completely dominated by inter-band Andreev processes and fixed to $2\sigma_0$.
For finite energies, the normalized conductance features a sharp change at $E\!=\!\varepsilon$, which determines the energy separation between Dirac cones in the dispersion relation.
The differential conductance of a SL-S junction and its derivative thus provide a very sensitive tool to study the creation of DPs by a superlattice potential.

\begin{figure}
\includegraphics[width=1\columnwidth]{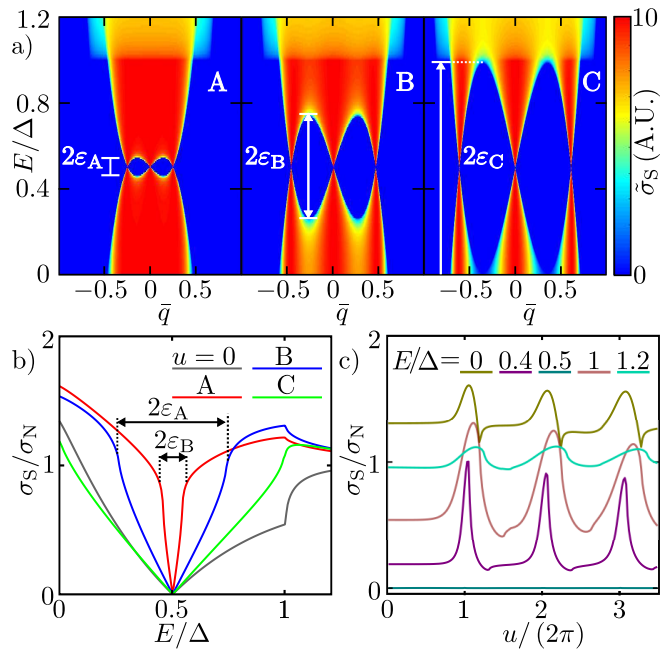}
\caption{Doped symmetric graphene superlattice.
a) Spectral differential conductance with $E_F \!=\! \Delta/2$ for different values of $u/\left(2\pi\right)\!=\!1.03$(A), $1.11$(B), $1.19$(C). In all maps $\hbar v_F q_\text{max}\!\simeq\!16\Delta$.
b) Differential conductance as a function of the energy for the different values of $u$ in a). The arrows indicate the parameter $\protect\varepsilon $.
c) Differential conductance as a function of $u$ for different energies.
In all cases $L\!=\!\protect\xi _{0}/2$. }
\label{fig:asym}
\end{figure}

\section{Non-ideal superlattice}

We now consider three deviations form the ideal superlattice potential described above. Namely, a finite doping on the graphene layer $E_F\!\neq\!0$, an asymmetry in the superlattice potential, or a SL region of finite length.

\subsection{Electrostatic asymmetry}
We start analyzing the effect of an asymmetry on the superlattice potential.
First, we consider the effect of a finite doping on the graphene layer with a perfectly symmetric SL potential with $U_n\!=\!U_p$ and $W_n\!=\!W_p$. Doping the graphene layer with $E_F\!\neq\!0$ shifts the position where the new DPs are created, cf. \fref{fig:asym}(a), but it does not change the condition for the their formation, $u\!=\!2n\pi$.
If $E_F\!<\!\Delta$, shifting the position of the DP to finite energies results in an enhanced suppression of the conductance, see \fref{fig:asym}(b,c), due to the vanishing density of states for hole-like excitations~\cite{Specular,Burset_2008}.
This is a unique property of graphene's gapless spectrum which, interestingly, is not affected by the creation of new Dirac points, in contrast to the undoped case with $E_F\!=\!0$ [compare \fref{fig:sym}(d) and \fref{fig:asym}(c)].
Since the splitting of the energy band into several DPs now takes place completely in the positive energy range (if $E_F\!>\!0$), the range of energies where normal reflections are enhanced is now $2\varepsilon$.
As before, an analysis of the derivative of the conductance allows us to estimate the value of $\varepsilon$, corresponding to the change of slope in the conductance.
A finite doping of the graphene layer, in the regime where inter-band Andreev reflections are enhanced, thus helps visualizing the formation of DPs using the normalized conductance.

An asymmetry in the superlattice potential has an important effect on the formation of new DPs. The asymmetry may be induced by changing the relative height ($U_p\!\neq\!U_n$) or width ($W_p\!\neq\!W_n$) of the potential barriers and wells. We can thus parametrize it defining $\alpha\!=\!W_n/W_p$ and $\beta\!=\!U_n/U_p$.
For an asymmetric superlattice, it is useful to define the average potential as the integral over a period~\cite{Arovas_2010}, resulting in $E_F^*\!=\!E_F+\omega$, with $\omega\!=\!U_p(\alpha\beta-1)/(1+\alpha)$. The position of the original DP under an asymmetric superlattice is given by $E_F^*$, so we henceforth refer to it as the \textit{effective} Fermi energy.
It is now possible to redefine the potential in \eref{eq:potencial} as
\begin{equation}\label{eq:potencial2}
V(x)\!=\!\left\{\!
\begin{array}{cl}
E_F^*+\omega\frac{1+\beta}{\alpha\beta-1}, & x\!\in\! \left[ mL,mL-W_{n}\right]  \\
E_F^*-\omega\frac{\alpha+\alpha\beta}{\alpha\beta-1}, & x\!\in\! \left[mL-W_{n},\left( m-1\right)L\right]
\end{array} \right. \!\!.
\end{equation}

\begin{figure}
\includegraphics[width=1\columnwidth]{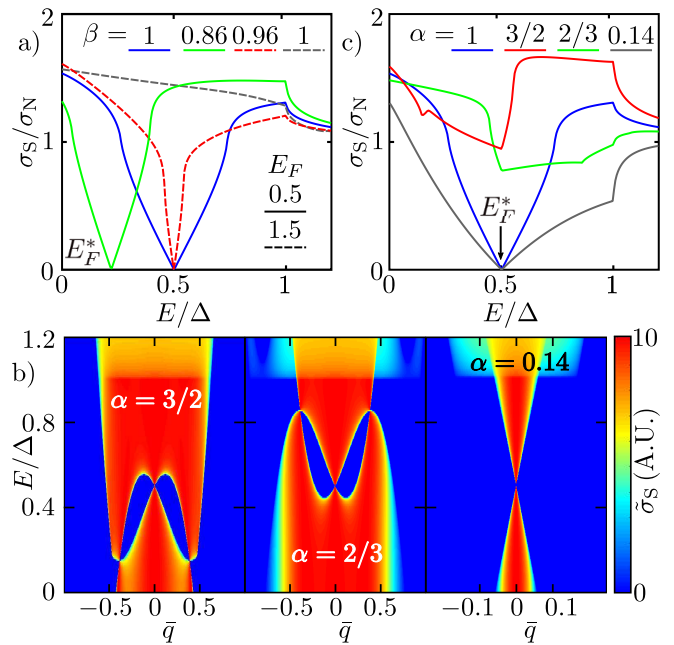}
\caption{Asymmetric superlattice potential characterized by parameters $\alpha$ and $\beta$ as defined in the text.
a) Asymmetry in the height of the potential barriers and wells: for $\alpha\!=\!1$ and $u/(2\pi)\!=\!1.11$, differential conductance with different $E^*_F$ and $\beta$. Dotted lines have $E_F\!=\!\Delta/2$ and solid lines $E_F\!=\!3\Delta/2$.
b) Asymmetry in the width of the potential barriers and wells: for $\beta\!=\!1$ and $u/(2\pi)\!=\!1.11$, differential conductance for different values of $\alpha$, adjusting $E_F$ so that $E^*_F\!=\!\Delta/2$ in all cases.
c) Spectral differential conductance for the red (left), green (center) and gray (right) lines in b). In all cases, $L\!=\!\xi_0/2$, and $\hbar v_F q_\text{max}\!\simeq\!15\Delta$ for the maps.
}
\label{fig:asym2}
\end{figure}

For the case where the potential barriers and wells have the same width ($\alpha\!=\!1$) but different heights ($\beta\!\neq\!1$), following the sketch of \fref{fig:potencial}, the barriers and wells take the values $E_F^*\pm\mean{V}$, respectively, with $\mean{V}\!=\!U_p(1+\beta)/2$.
The main effect of the asymmetry is to shift the Fermi energy to the effective one, $E_F^*$, where the original Dirac cone is and the new DPs are created, cf. \fref{fig:asym2}(a).
For reference, we also plot in \fref{fig:asym2}(a) a case with a symmetric SL potential taken from \fref{fig:asym}(b) (blue line).
By changing the value of $\beta$, the conductance features a dip at different energies below the gap, as long as $E_F^*\!<\!\Delta$ (green line).
If the asymmetry is such that $E_F^*\!>\!\Delta$, the conductance does not exhibit any dip and approaches the case of a heavily-doped graphene layer, even if $E_F\!<\!\Delta$ (dashed gray line).
The doping $E_F$ of the graphene layer thus provides an experimentally controllable parameter that can rectify any asymmetry in the height of the barriers and wells of the SL potential.

\subsection{Spatial asymmetry}
When the asymmetry on the SL potential affects the widths of the barriers and wells ($\alpha\!\neq\!1$), the impact on the conductance is more pronounced.
In this situation, the position of the original DP is still given by $E_F^*$, but the new DPs appear at different energies.
For example, the real doping $E_F$ in the three panels of \fref{fig:asym2}(b) has been adjusted so that all cases have $E_F^*\!=\!\Delta/2$.
However, the barrier's width is bigger than the corresponding for the wells in the left panel ($\alpha\!>\!1$), while it is smaller in the other panels ($\alpha\!<\!1$). As a result, the new DPs are created for energies below or above the effective Fermi energy $E_F^*$, respectively.
In the right panel, the superlattice is so asymmetric that the new DPs merge back into the original Dirac cone, recovering the result for a heavily-doped graphene layer.

The spatial asymmetry drastically changes the conductance, as we show in \fref{fig:asym2}(c). As a reference, we show a symmetric result with finite doping $E_F\!=\!\Delta/2$ (blue line) and compare it to the asymmetric cases from \fref{fig:asym2}(b), where the doping was adjusted so that $E_F^*\!=\!\Delta/2$.
When $\alpha\!<\!1$ (green line), the new DPs appear for energies bigger than the effective Fermi energy and their impact on the conductance is diminished.
On the other hand, if $\alpha\!>\!1$ (red line), the new DPs appear for smaller energies and can be clearly seen as kinks in the conductance.
The case with a high asymmetry (gray line) approximates the conductance of a doped graphene layer with dominant intra-band Andreev reflection~\cite{Specular}.

Our results thus show the importance of a regularly spaced superlattice potential, while the asymmetry in the electrostatic barriers can be compensated by an uniform change in the doping level of the graphene layer.

\begin{figure}
\includegraphics[width=1.\columnwidth]{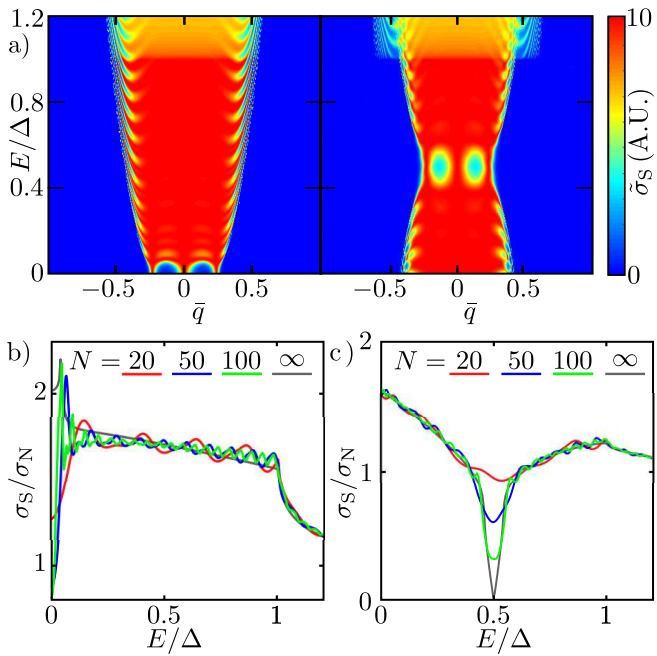}
\caption{Finite superlattice potential.
(a) Spectral differential conductance for $N\!=\!50$, with $E_F\!=\!0$ (left) and $E_F\!=\!\Delta/2$ (right). In both cases $\hbar v_F q_\text{max}\!\simeq\! 14\Delta$.
(b,c) Differential conductance for different number of n-p junctions $N$ of the finite superlattice potential, with $E_F\!=\!0$ (b) and $E_F\!=\!\Delta/2$ (c). The gray line recovers the semi-infinite superlattice potentials in \fref{fig:sym} and \fref{fig:asym}. In all cases, $L\!=\!\xi_0/2$.
}
\label{fig:finite}
\end{figure}

\subsection{Finite-length superlattice}
The emergence of new Dirac points and their separation from the original cone in a semi-infinite SL potential can be discerned in the differential conductance by the parameter $\varepsilon$.
We now analyze the impact of a finite length SL potential on the previous results.
We consider a finite-size graphene SL contacted on one side ($x\!=\!0$) to a superconductor, and on the other side ($x\!=\!-NL$) to a normal state reservoir, which we model as a heavily-doped graphene semi-infinite layer~\cite{Burset_2008,Herrera_2010,Oscar_2018}. Here, $L$ is the size of a n-p junction and $N$ the total number of n-p junctions in the finite SL region.
Nano-scale hybrid junctions where the reservoirs and the intermediate scattering region are built from different materials present very different interface transmissions~\cite{Klapwijk_Review,Jonas_2017}. For simplicity, we only consider here transparent couplings between the finite SL and the normal and superconducting leads and a symmetric SL potential.

The finite length of the intermediate region results in the splitting of the continuous Dirac cone into energy bands, cf. \fref{fig:finite}(a).
In the presence of a SL potential, the condition for the emergence of new DPs, $u=2n\pi$, is roughly maintained even for a very small SL.
A \textit{menorah}-like pattern similar to that of \fref{fig:sym}(b) emerges for small lengths, although the resonances at the positions of the DPs are broadened and not so well defined.
The broadening of the dispersion relation for subgap energies is shown in \fref{fig:finite}(a) for the undoped (left panel) and doped cases (right panel). The blurring of the DPs results in an increased probability of normal backscattering.
We show the finite-length effect in the differential conductance in \fref{fig:finite}(b) and (c) for $E_F\!=\!0$ and $\Delta/2$, respectively.
For reference, the gray lines show the behavior of a semi-infinite superlattice potential with the same parameters.
We confirm the robustness of the effect of the SL in the differential conductance for all energies different than $E_F$.
At the charge neutrality point $E_F$, the semi-infinite case is qualitatively reproduced for superlattice lengths $N\!\gtrsim\!50$, with $L\!=\!\xi_0/2$.
However, to estimate the value of $2\varepsilon$ for a doped superlattice, only a few p-n junctions ($N\!\sim\!20$) are needed.


\section{Conclusions}
We have analyzed the transport properties of a graphene SL-S junction. We have demonstrated that the superlattice potential can create new DPs even in the presence of superconducting correlations.
Moreover, the changes that the SL potential causes on the graphene's spectrum are enhanced for subgap energies.
Therefore, the emergence of new DPs can be easily monitored through the differential conductance of the junction.
The normalized conductance features sharp changes for energy values equal to a parameter, $\varepsilon$, determined by the separation between newly created DPs and the original cone.
Further, the superconducting conductance is always enhanced over the normal state one for energies below but close to the gap $\Delta$.
These effects are robust in the presence of asymmetry in the superlattice potential and finite-size, as long as the formation of new DPs is possible.
Our results thus suggest that superconducting hybrid junctions are useful to experimentally observe the modification of the band dispersion relation due to a superlattice potential on a graphene layer.



\acknowledgments
We acknowledge funding from DIEB, Universidad Nacional de Colombia  project No. 34916, the Horizon 2020 research and innovation programme under the Marie Sk\l odowska-Curie Grant No. 743884, Academy of Finland (project No. 312299), and Spanish MINECO via grants FIS2015-74472-JIN (AEI/FEDER/UE), FIS2014-55486P, FIS2017-84860-R and through ``Mar\'\i a de Maeztu" Programme for Units of Excellence in R\&D (MDM-2014-0377).

\appendix

\section{Superlattice Green function. \label{sec:app1}}

To analyze the transport properties of the SL-S junction, we calculate the Green function of the system.
The building block for a semi-infinite graphene in the  superlattice space is the Green function of an isolated zigzag graphene layer finite in the $x$-direction:
\begin{gather}
\hat{g}_{0}^{<(>)}(x,x^{\prime }) = \frac{-i}{2\hbar v_{F}\left( 1-r_{L}^{B}r_{R}^{A}\right) \cos \alpha }
\notag \\
\times \left( e^{ik_{x} \left|x^{\prime }-x\right|} \hat{f}_{\mp}
+ r_{L}^{B}r_{R}^{A} e^{- ik_{x}\left| x^{\prime } - x\right| } \hat{f}_{\pm} \right. \notag \\
\left.
+ r_{L}^{B} e^{ ik_{x}\left( x+x^{\prime}\right)} h_{+}
+ r_{R}^{A} e^{-ik_{x}\left( x+x^{\prime }\right) } h_{-} \right)  ,  \label{eq:fdg-fin}
\end{gather}
with
\begin{equation*}
\hat{f}_{\pm}=\!\left(\!
\begin{array}{cc}
1 & \pm e^{\pm (-i\alpha )} \\
\pm e^{\pm i\alpha } & 1%
\end{array}\!
\right) , \text{ }\hat{h}_{\pm}=\! \left(\!
\begin{array}{cc}
e^{\pm (-i\alpha )} & \mp 1 \\
\pm 1 & -e^{\pm i\alpha }%
\end{array}\!
\right) , \text{\ }
\end{equation*}
and
\begin{eqnarray*}
r_{L}^{A} &=&-e^{i\alpha }e^{-2ik_{x}x_{L}},\text{ \ }r_{R}^{A}=-e^{-i\alpha
}e^{2ik_{x}x_{R}}, \\
r_{L}^{B} &=&e^{-i\alpha }e^{-2ik_{x}x_{L}},\text{ \ }r_{R}^{B}=e^{i\alpha
}e^{2ik_{x}x_{R}}.
\end{eqnarray*}
Here,  $v_F$ is the Fermi
energy and velocity, respectively, of the graphene sheet. The symbol $<(>)$ indicates $x<x^{\prime }$ $%
(x>x^{\prime })$. The explicit calculation of \eref{eq:fdg-fin} is given in
Ref.~\onlinecite{Herrera_2010}.

One superlattice period (n-p junction) is composed of two finite graphene regions with different Fermi energies, and widths $W_{p,n}$. The Green function for each region is given by \eref{eq:fdg-fin} and they can be coupled using Dyson's equation to obtain the Green function of the finite n-p junction. Assuming that the coupled region extends from $x_L=-L$ to $x_R=0^-$, we obtain the Green functions
\begin{subequations} \label{g4}
\begin{align}
\hat{g}_{np}^{>}\left( -L,-L\right) ={}& \frac{-i}{\hbar v_{F}}\left(
\begin{array}{cc}
-C_{p} & 0 \\
1 & 0%
\end{array}%
\right) ,  \label{g1} \\
\hat{g}_{np}^{<}(-L,0^-)={}& \frac{-i}{\hbar v_{F}}\left(
\begin{array}{cc}
0 & -D_{pn} \\
0 & 0%
\end{array}%
\right) ,  \label{g2} \\
\hat{g}_{np}^{<}\left( 0^-,0^-\right) ={}& \frac{-i}{\hbar v_{F}}\left(
\begin{array}{cc}
0 & -1 \\
0 & -C_{n}%
\end{array}%
\right) ,  \label{g3} \\
\hat{g}_{np}^{>}\left( 0^-,-L\right) ={}& \frac{-i}{\hbar v_{F}}\left(
\begin{array}{cc}
0 & 0 \\
-D_{pn} & 0%
\end{array}%
\right) .
\end{align}
\end{subequations}

The Green function for a finite superlattice of length $NL$, with $L$ the period of a n-p junction, is obtained coupling \eref{g4} to each other $N$ times.
The resulting Green function reads
\begin{align}
\tilde{g}_{fsl}={}& \frac{-i}{\hbar v_F}\left(\!
\begin{array}{cc}
-C^N_n & 0 \\
1 & 0%
\end{array}\!
\right) , \label{eq:finite-gf} \\
C^N_n={}& \frac{C^{N-1}_n t^2(C^{N-1}_p - D^{N-1}_{pn})+C^{N-1}_p}{C^{N-1}_n C^{N-1}_p t^2+1} . \notag
\end{align}

The Green function of the semi-infinite superlattice is calculated using the self-similarity of a semi-infinite chain.
Adding one block to a semi-infinite number of graphene n-p blocks results in the same semi-infinite chain. Therefore, it is possible to analytically calculate the Green function for the complete superlattice evaluated at one edge, $\hat{g}_\text{SL}(0^-,0^-)$, by imposing that it is equal the Green function of the superlattice when a
new block $\hat{g}_{np}$ has been added. Using Dyson's equation we find
\begin{gather}
\hat{g}_{\text{SL}}(0^-,0^-) =\hat{g}_{np}(0^-,0^-) \label{gsl} \\
-\hat{g}_{np}(0^-,-L)\hat{\Sigma}^{\dagger} \hat{M}^{-1}\hat{g}_{\text{SL}}(-L,-L)\hat{\Sigma}\hat{g}_{np}(-L,0^-),  \notag
\end{gather}
with
\begin{align*}
\hat{M}={}& I - \hat{g}_\text{SL}(-L,-L) \hat{\Sigma} \hat{g}_{np}(-L,-L)\hat{\Sigma}^{\dagger} , \\
\hat{\Sigma} ={}& \tilde{t}\left(
\begin{array}{cc}
0 & 1 \\
0 & 0%
\end{array}%
\right) .
\end{align*}
Solving \eref{gsl}, we obtain \eref{eq:gf_SL} of the main text.

\section{Differential conductance for the superlattice--superconductor junction. \label{sec:app2}}

Once the Green function for the coupled SL-S system is obtained, the electric current follows from Keldysh formalism. Following the extension of
the Hamiltonian approach described in Refs.~\onlinecite{Cuevas_1996,Oscar_2018}, the zero-temperature differential conductance reads as
\begin{widetext}
\begin{align}
\tilde{\sigma} _\text{Q1}={}&t^{2}\text{Tr}\left[ \text{Re}\left( \left( \hat{I}+t%
\hat{G}_{t,ee}^{r}\sigma _{1}^{T}\right) \hat{\rho}_{sc,ee} \left( \hat{I}+t\sigma _{1}\hat{G}_{t,ee}^{r\ast }\right) \bar{%
\rho}_{sl,e}\right) \right] +t^{4}\text{Tr}\left[ \text{Re}\left( \hat{G}%
_{t,eh}^{r\ast }\bar{\rho}_{sl,e}\hat{G}_{t,eh}^{r}\sigma _{1}^{T}%
\hat{\rho}_{sc,hh} \sigma _{1}\right) \right], \\
\tilde{\sigma} _\text{Q2} ={}&-t^{3}\text{Tr}\left[ \text{Re}\left( \left( \hat{I}
+ t\sigma _{1}\hat{G}_{t,ee}^{r\ast }\right) \bar{\rho}_{sl,e}\hat{G}
_{t,eh}^{r}\sigma _{1}^{T}\hat{\rho}_{sc,he}
+
\left( \hat{I}+t\hat{G}_{t,ee}^{r}(E)\sigma
_{1}^{T}\right) \hat{\rho}_{sc,eh} \sigma _{1}\hat{G}%
_{t,eh}^{r\ast }\bar{\rho}_{sl,e}\right) \right] , \\
\tilde{\sigma} _\text{A} ={}&t^{4}\text{Tr}\left[ \text{Re}\left( \hat{G}%
_{sc,eh}^{r\ast }\bar{\rho}_{sl,e}\hat{G}_{sc,eh}^{r}\bar{\rho}%
_{sl,h}+\hat{G}_{sc,he}^{r\ast }\bar{\rho}_{sl,h}\hat{G}%
_{sc,he}^{r}\bar{\rho}_{sl,e}\right) \right] ,  \label{andzz}
\end{align}
\end{widetext}
with $\bar{\rho}_{sl,e(h)}\!=\!\hat{\sigma}_{1}\hat{\rho}_{sl,ee(hh)}
\hat{\sigma}_{1}^{T}$.
The density of states $\rho _{sl}$ and $\rho _{sc}$ are related to the Green functions of the decoupled system at equilibrium, and are defined as $\rho (E)\!=\!\frac{i}{2\pi }\left( g-g^{\dagger}\right)$, where $sl$ denote the superlattice and $sc$ the superconductor at $x\!=\!0 $.

The Green functions of the coupled junction (SL-S), $\hat{G}_{sc,eh(he)}^{r}$ and $\hat{G}_{t,ee(eh)}^{r}$, are matrices in graphene sublattice space, representing elements on the Nambu space.
The coupling is realized via Dyson's equation as follows:
\begin{align*}
 \tilde{G}_{sc}^{r}={}& \tilde{g}_{sc}^{r}\left(
0^{+},0^{+}\right) + \hat{G}_{t}^{r} \tilde{\Sigma} \tilde{g}_{sc}^{r}\left( 0^{+},0^{+}\right) , \\
\tilde{G}_{t}^{r} ={}& \tilde{g}_{sc}^{r}(0^{+},0^{+}) \tilde{\Sigma}^{\dagger } \tilde{P}^{-1}\tilde{g}_{\text{SL}}^{r}(0^{-},0^{-}) , \\
\tilde{P} ={}& \hat{I}-\tilde{g}_{\text{SL}}(0^{-},0^{-})\tilde{\Sigma}\tilde{g}_{sc}(0^{+},0^{+}) \tilde{\Sigma}^{\dagger} ,
\end{align*}
where $\tilde{g}_{sc}^{r}(0^{+},0^{+}) $
corresponds to the Green function in equilibrium for the superconducting region and is given in Ref.~\onlinecite{Herrera_2010}.

In the case of a finite superlattice, the differential conductance is also given by \eref{conductance}, substituting the Green function for the semi-infinite superlattice by a new function representing the coupling of semi-infinite, heavily-doped graphene contact, $\tilde{g}_{nc}$, cf. Refs.~\onlinecite{Burset_2008,Herrera_2010}, with the finite superlattice $\tilde{g}_{fsl}$, \eref{eq:finite-gf}.







\bibliography{bib_superlattice}

\end{document}